\documentclass[a4paper]{article}

\usepackage{INTERSPEECH2021}
\usepackage{multirow}
\usepackage{url}
\usepackage{amsmath}
\usepackage{color}
\usepackage{cite}
\usepackage[dvips]{epsfig}

\DeclareMathOperator*{\argmax}{arg\,max}

\title{Phrase break prediction with bidirectional encoder representations in Japanese text-to-speech synthesis}
\name{Kosuke Futamata, Byeongseon Park, Ryuichi Yamamoto, Kentaro Tachibana}
\address{LINE Corp., Tokyo, Japan}
\email{\{kosuke.futamata, park.byeongseon, ryuichi.yamamoto, kentaro.tachibana\}@linecorp.com}

\begin{document}

\maketitle

\begin{abstract}
We propose a novel phrase break prediction method that combines implicit features extracted from a pre-trained large language model, a.k.a BERT, and explicit features extracted from BiLSTM with linguistic features.
In conventional BiLSTM-based methods, word representations and/or sentence representations are used as independent components.
The proposed method takes account of both representations to extract the latent semantics, which cannot be captured by previous methods.
The objective evaluation results show that the proposed method obtains an absolute improvement of 3.2 points for the F1 score compared with BiLSTM-based conventional methods using linguistic features.
Moreover, the perceptual listening test results verify that a TTS system that applied our proposed method achieved a mean opinion score of 4.39 in prosody naturalness, which is highly competitive with the score of 4.37 for synthesized speech with ground-truth phrase breaks.


\end{abstract}

\section{Introduction}
Deep learning-based end-to-end approaches have achieved great success in text-to-speech (TTS) synthesis.
In particular, TTS systems based on sequence-to-sequence models (e.g., Tacotron\cite{tacotron,tacotron2}) enable the models to directly map character sequences to acoustic features, thereby eliminating the need for a complicated text processing front-end.
However, front-end modules, such as language-dependent text normalization and grapheme-to-phoneme (G2P) conversion, are still beneficial as pre-processing steps, and can be crucial for practical applications\cite{li2019close,kastner2019representation}.
Furthermore, in addition to text normalization and G2P conversion, prosodic features, including an accent phrase and phrase break, also play an important role in pitch-accent languages like Japanese\cite{Fujimoto2019, kurihara2021}.

The front-end of a TTS system depends on the language.
For Japanese, it includes a variety of modules, such as text normalization,  accent estimation\cite{TTS_features_accent}, G2P conversion\cite{TTS_features_g2p_1, TTS_features_g2p_2}, and phrase break prediction\cite{pbp_random_forest_1, TTS_features_break}.
In this study, we focus on Japanese phrase break prediction, which is one of the essential tasks for improving the naturalness of the prosody.

Previous studies can be divided into two categories.
One is traditional statistical methods\cite{pbp_random_forest_1, pbp_crf_en_1, pbp_cart_ja_1} using manually designed linguistic features.
The other category includes deep learning-based sequential models, such as Reccurent Neural Network (RNN)\cite{pbp_rnn_1, pbp_rnn_2, pbp_lstm_1}, and Long-Short Term Memory (LSTM)\cite{pbp_lstm_1, pbp_lstm_2}.
These methods achieve much better performance than traditional statistical methods.
However, the main problem with these methods is that they usually require a large amount of labeled data to achieve good performance.
Therefore, conventional methods have made use of unsupervised pre-trained representations, including word representations and/or sentence representations \cite{pbp_lstm_1, pbp_lstm_2, pbp_amazon, pbp_word_embedding} to improve performance, even with small data.
Although these unsupervised representations are helpful, their performance is potentially limited because word representations and sentence representations are typically extracted by separate modules without modeling complex semantic structures.

To address these issues of previous methods, inspired by the great success of Bidirectional Encoder Representations from Transformers (BERT)\cite{bert}, which is an unsupervised pre-trained large language model applied in many NLP tasks, we propose a novel method that uses both labeled and unlabeled data for phrase break prediction.
The proposed method uses both explicit features from LSTM using various linguistic features and implicit features from BERT to model the relations between features and phrase break information.
Because BERT can learn latent word and sentence representations with semantic meanings simultaneously, it provides a better representation than simple word or sentence embedding representations.
Additionally, because the model can use explicit features extracted from LSTM using various linguistic features as input, it makes it possible for the model to improve performance, even for a limited amount of training data.

We compared the proposed method in both an objective evaluation based on the F1 score and subjective evaluations based on the mean opinion score (MOS) test and AB preference test.
The results of the objective evaluation showed that the proposed method helped the model to achieve a significant improvement of 3.2 points for the F1 score compared with a BiLSTM-based conventional method using linguistic features.
Moreover, the results of the subjective evaluations verified that the TTS system that applied our proposed method achieved an MOS of 4.39 in terms of prosody naturalness, which is highly competitive with the score of 4.37 for synthesized speech with ground-truth phrase breaks.
Additionally, there were significant differences between the proposed method and the conventional methods for the AB preference test.

\section{Phrase break prediction in TTS front-end}
A phrase break is defined as a phonetic pause inserted between phrases, and it occurs because of breathing and/or an intonational reset.
In a text, phrase breaks are usually represented as a type of punctuation.
When the TTS system observes a comma in English text, it is generally accepted that a phrase break should be inserted, as the author intends the words before and after the punctuation to be separated.
However, this rule-based approach usually does not work well.
It has been reported that the rule-based approach results in about half of the phrase breaks being inserted correctly, and a few breaks are inserted at incorrect positions\cite{taylor2009}.
These errors come from the fact that human speakers usually insert phrase breaks without punctuation, for example, for breathing, expression, and accent change.

The main goal of phrase break prediction is to predict the positions of superficial and latent phrase breaks from the input text.
This task can be formulated as sequential labeling, which labels a phrase break or non-phrase break for each token after applying some text pre-processing modules, such as text normalization and tokenization;
that is, given a source sequence $X = \{x_1, x_2, ..., x_t\}$ and label sequence $Y = \{y_1, y_2, ..., y_t\}$ corresponding to a source sequence $X$, the goal is to predict a label sequence $Y$ from a token sequence $X$, and the $i$-th element in label sequence $Y$ is defined as a binary label with a non-phrase break ($y_i$ = 0) or phrase break ($y_i$ = 1).
Given the model parameter $\theta$ and source sequence $X$, we aim to calculate parameters $\hat{\theta}$ by maximizing the log-likelihood of the given label sequence $Y$ as follows:

\begin{equation}
    \hat{\theta} = \argmax_{\theta} \sum_{i = 1}^{t} \log p ( Y^{(i)} | X^{(i)} ; \theta )
    \label{equation:mle}
\end{equation}

The deep learning-based model architecture has been successfully applied to phrase break prediction.
Applications of bidirectional LSTM (BiLSTM) have been studied as sequence labeling tasks\cite{pbp_amazon, pbp_lstm_1, pbp_lstm_2, pbp_word_embedding}.
These methods use part-of-speech (POS) tags, dependency tree features, pre-trained word embedding, and character embedding in addition to the input sequences to improve performance.
Using BiLSTM enables the model to learn long-term dependencies from both the left and right direction, and add a variety of linguistic features as the input enriches contextual information in long-form dependencies.
In previous studies, not only a source sequence but also these linguistic features were used to predict phrase breaks.

\section{Proposed method using bidirectional encoder representations}
The main problem with the TTS front-end is that it costs a great deal to prepare a large labeled corpus.
Therefore, in conventional approaches, unsupervised pre-training methods, such as word embedding and sentence representations using a BiLSTM-based language model, are used to improve performance.
However, these representations are weak context representations; pre-trained word embedding contains semantic information for only a single word without sequential dependencies, and the BiLSTM language model lacks sufficient contextual information for a much longer sequence, although it learns long-term dependencies more easily than vanilla RNN.
To solve these problems, BERT\cite{bert} model, which is a recently proposed NLP pre-training method, is widely used to capture the long-term context dependency in sequences.
In this section, we briefly explain the general architecture and properties of BERT model, in addition to the proposed method that combines BiLSTM and BERT for phrase break prediction.

\subsection{BERT architecture}
BERT is a recently proposed language model composed of bi-directional hierarchical Transformer\cite{transformer} blocks.
The model can be used as an encoder that takes a series of subword tokens as input and generates a word embedding for each token.
The multi-head self-attention mechanism in Transformer blocks enables the model to capture word dependencies for both the left and right side context without any restriction on the position of words in a sentence.
BERT is fine-tuned for each task after two unsupervised pre-training tasks called masked language modeling (MLM), as well as next sentence prediction (NSP).
Benefit from this pre-training, the model is assumed to be capable of capturing rich contextual and semantic information of Japanese language, thereby facilitating the downstream NLP tasks (i.e., Phrase break prediction in this work).

\begin{figure}[tb]
  \begin{center}
  \epsfig{figure=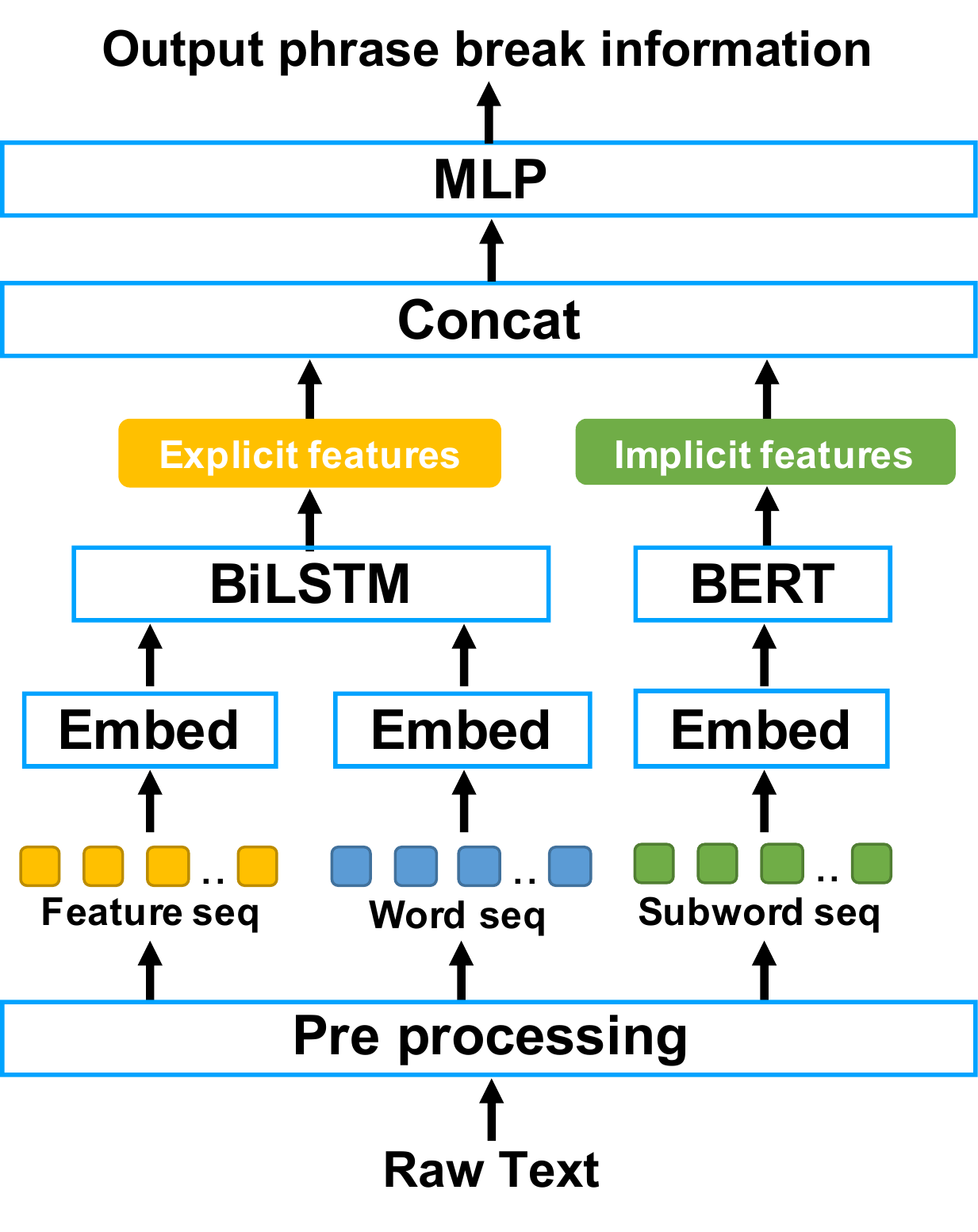,width=0.79\linewidth}
  \end{center}
  \caption{Architecture of the proposed method using BiLSTM with BERT for phrase break prediction}
  \label{fig:pbp_architecture}
\end{figure}

\begin{table*}[t]
    \centering
    \caption{Performance of different systems on Japanese for the phrase break prediction task}
    \scalebox{0.88}{
    \begin{tabular}{c||c|c|c|c|c|c|c|c}                                                                                                               \hline
       System   & Model                               & Input features           & \begin{tabular}{c} True \\ positives \end{tabular} & \begin{tabular}{c} False \\ negatives \end{tabular} & \begin{tabular}{c}False \\ positives \end{tabular} & F1               & Precision           &  Recall           \\ \hline
       1        & BiLSTM (Tokens)                     & Tokens                   &   653           &  114            &   53            & 88.7            & 92.5               & 85.1             \\ \hline
       2        & BiLSTM (Features)                   & Tokens+POS+DEP+W2V       &   676           &  91             &   56            & 90.2            & 92.3               & 88.1             \\ \hline
       3        &  BERT                               & Tokens                   &   688           &  79             &   39            & 92.1            & \textbf{94.6}      & 89.7             \\ \hline
       4        &  BiLSTM (Tokens) + BERT             & Tokens                   &   690           &  77             &   42            & 92.1            & 94.3               & 90.0             \\ \hline
       \textbf{5}        &  \textbf{BiLSTM (Features) + BERT}  & Tokens+POS+DEP+W2V       &   709           &  58             &   43            & \textbf{93.4}   & 94.3               & \textbf{92.4}    \\ \hline
    \end{tabular}
    }
    \label{tab:objective_evaluation}
\end{table*}

\begin{table*}[t]
    \centering
    \caption{Performance of different systems on Japanese for phrase break prediction with and without punctuation}
    \scalebox{0.8}{
    \begin{tabular}{c||c|c|c|c|c|c||c|c|c|c|c|c} \hline
    \multirow{2}{*}{System} & \multicolumn{6}{c||}{With punctuation}   & \multicolumn{6}{c}{Without punctuation}  \\ \cline{2-13} 
    & \begin{tabular}{c} True\\ positives\end{tabular} & \begin{tabular}{c} False\\ negatives\end{tabular} & \begin{tabular}{c} False\\ positives\end{tabular} & F1 & Precision & Recall & \begin{tabular}{c} True\\ positives\end{tabular} & \begin{tabular}{c} False\\ negatives\end{tabular} & \begin{tabular}{c} False\\ positives\end{tabular} & F1 & Precision & Recall \\ \hline\hline
1   &  357  & 0 & 1 & 99.9 & 99.7 & 100  & 296 & 114 & 52 & 78.1          & 85.1          & 72.2 \\ \hline
2   &  357  & 0 & 1 & 99.9 & 99.7 & 100  & 319 & 91  & 55 & 81.4          & 85.3          & 77.8 \\ \hline
3   &  355  & 2 & 0 & 99.7 & 100  & 99.4 & 333 & 77  & 39 & 85.2          & \textbf{89.5} & 81.2 \\ \hline
4   &  356  & 1 & 1 & 99.7 & 99.7 & 99.7 & 334 & 76  & 41 & 85.1          & 89.1          & 81.5 \\ \hline
\textbf{5}   &  357  & 0 & 1 & 99.9 & 99.7 & 100  & 352 & 58  & 42 & \textbf{87.6} & 89.3          & \textbf{85.9} \\ \hline
\end{tabular}
}
\label{tab:objective_evaluation_detail}
\end{table*}

\subsection{Application to phrase break prediction}
The proposed method predicts phrase break positions using implicit features extracted from BERT and explicit features extracted from the conventional method, BiLSTM.
Figure \ref{fig:pbp_architecture} shows the architecture of our proposed method.

The input Japanese text is firstly pre-processed by text normalization, tokenization, POS tagging, and dependency parsing modules.
Explicit and implicit features are then extracted by BERT and BiLSTM encoder.
On the BiLSTM side, the word sequence, POS tags, and dependency tree features corresponding to the sequence are fed into the model, and then explicit features are generated from the last layer.
Word embedding in BiLSTM is pre-trained on Japanese Wikipedia.
Conversely, on the BERT side, a subword sequence is fed into the model, and then implicit features are generated from a weighted average of all the layers.
We use a weighted average of all the layers instead of using only the last layer of BERT, which is generally used for NLP downstream tasks, because different BERT layers have different types of information\cite{bertology}.
The lower layers have the most linear word order information, the middle layers have the most syntactic information, and the final layer has the most task-specific information.
Preliminary experiments showed that using a weighted average of all the layers contributed much more to performance than using only the last layer.

Explicit features from BiLSTM and implicit features from BERT are finally concatenated, and then the method determines whether a phrase break should be labeled after each token.
We expect that implicit features extracted from BERT will learn the rich syntactic and semantic information of each word in a given context, which cannot be captured by only adding the BiLSTM layer.
Conversely, we expect that explicit features extracted from BiLSTM should capture the long-term dependencies in a source sequence directly, thereby helping the implicit features to be much more rich representations.

\begin{table}[t]
    \centering
    \caption{MOS results with 95\% confidence intervals}
    \scalebox{1.0}{
    \begin{tabular}{c||c|c}                                                            \hline
       System    & Model                                                & MOS                           \\ \hline
       1         & Reference (Natural)                                  &  $4.70 \pm{0.05}$             \\ \hline
       2         & Reference (TTS)                                      &  $4.37 \pm{0.05}$             \\ \hline\hline
       3         & Rule-based                                           &  $3.62 \pm{0.07}$             \\ \hline
       4         & BiLSTM (Tokens)                                      &  $3.80 \pm{0.07}$             \\ \hline
       5         & BiLSTM (Features)                                    &  $4.05 \pm{0.06}$             \\ \hline
       6         & BERT                                                 &  $4.26 \pm{0.05}$             \\ \hline
       \textbf{7}         & \textbf{BiLSTM (Features) + BERT}                             &  $\textbf{4.39} \pm{0.05}$    \\ \hline
    \end{tabular}
    }
    \label{tab:mos_evaluation_result}
\end{table}

\section{Objective evaluation}

\subsection{Experimental setting}
We compared the performance of the proposed method with different systems based on either BiLSTM or BERT using an objective evaluation.
For the dataset, we collected 99,907 utterances, where each utterance was transcribed and had silence between words, and word transitions with a silence of more than 200 ms were marked as phrase breaks.
We split the dataset into three subsets of 98,807, 500, and 500 sentences for training, validation, and test, respectively.

In the objective evaluation, we compared the following five methods.
(1) \textbf{BiLSTM (Tokens)}: conventional BiLSTM-based method using only a source sequence\cite{pbp_lstm_1}.
(2) \textbf{BiLSTM (Features)}: (1) BiLSTM (Tokens)-based method that adds designed linguistic features\cite{pbp_amazon}, including POS tags\cite{sudachi}, dependency tree features (DEP)\cite{ginza}, and pre-trained word embedding.
(3) \textbf{BERT}: method that uses BERT model.
(4) \textbf{BiLSTM (Tokens) + BERT}: method that combines (1) BiLSTM (Tokens) and (3) BERT.
(5) \textbf{BiLSTM (Features) + BERT}: proposed method that combines (2) BiLSTM (Features) and (3) BERT.
We set up baselines as conventional approaches; (1) BiLSTM (Tokens) and BiLSTM (Features).

For all experiments, we used a two-layer BiLSTM (with 512 dimensions).
We set the embedding sizes of linguistic features to 32.
The model was trained for 20 epochs with Adam optimizer\cite{adam} with a minibatch size of 64 utterances.
The learning rate was held constant at 1e-5.
The loss function that we used to train the model was cross-entropy loss, and we evaluated the model using the F1 score.
We stopped training when the F1 score did not increase for 10 epochs.
For BiLSTM (Features) and BiLSTM (Features) + BERT, we used word embedding vectors pre-trained on Japanese Wikipedia articles.
For BERT-based models, we used a BERT-Base model pre-trained on Japanese Wikipedia articles and released on Github\cite{cl-tohoku}.

\begin{table*}[htb]
    \centering
    \caption{AB preference results for the one-tailed binomial test, n.s.: no difference, ***: significant differences with $p \leq .001$}
    \scalebox{1.0}{
    \begin{tabular}{c|c||c|c|c|c}                                                                                                                                           \hline
       Model \ A                                           & Model \ B                                           &  Preference \ A & Preference \ B &  Neutral  & binomial test     \\ \hline\hline
       Rule-based                                           & BiLSTM (Tokens)                                      & 125             & \textbf{335}   & 410       &  ***   \\ \hline
       BiLSTM (Tokens)                                      & BiLSTM (Features)                                    & 172             & \textbf{279}   & 419       &  ***    \\ \hline
       BiLSTM (Features)                                    & BERT                                                 & 138             & \textbf{329}   & 403       &  ***    \\ \hline
       BERT                                                 & BiLSTM (Features) + BERT                             & 144             & \textbf{238}   & 488       &  ***    \\ \hline
       BERT                                                 & Reference (TTS)                                      & 156             & \textbf{269}   & 445       &  ***    \\ \hline
       BiLSTM (Features) + BERT                             & Reference (TTS)                                      & 151             & \textbf{156}   & 563       &  n.s.    \\ \hline
    \end{tabular}
    }
    \label{tab:ab_test_result}
\end{table*}

\subsection{Results and analysis}
The experimental results of different systems are presented in Table \ref{tab:objective_evaluation} and Table \ref{tab:objective_evaluation_detail}.
Table \ref{tab:objective_evaluation} shows all results regardless of whether a phrase break was provided by punctuation in the utterance, and Table \ref{tab:objective_evaluation_detail} shows separate results when a phrase break was provided with and without punctuation.

The experimental results presented in Table \ref{tab:objective_evaluation} show that the proposed method improved the F1 score compared with the conventional methods.
Comparing BERT-based models with the conventional methods, BiLSTM (Tokens) and BiLSTM (Features), the BERT-based models improved the F1 score significantly,
This implies that the proposed method achieved high performance, even for a limited amount of training data.
The proposed method achieved absolute improvements of 4.7 points and 3.2 points compared with BiLSTM (Tokens) and BiLSTM (Features), respectively, which indicates that implicit features extracted from BERT are much more effective than explicit features only used in the conventional methods.
Furthermore, the F1 score of the proposed method was much higher than that of BiLSTM (Tokens) + BERT, although the performances of BERT and BiLSMT (Tokens) + BERT were the same.
This implies that explicit features extracted from BiLSTM using various linguistic features were used effectively for predicting phrase breaks.

\section{Subjective Evaluation}

\subsection{Experimental setting}

\subsubsection{TTS setup}
To verify the effectiveness of the proposed phrase break prediction method in TTS scenarios, 
we combined the proposed method as part of the text processing front-end of a FastSpeech2 based TTS system \cite{fastspeech2}.
The TTS system consisted of two models: (1) a feed-forward Transformer-based acoustic model that predicts acoustic features from a phoneme sequence with additional phrase break information, and (2) a parallel WaveGAN vocoder that generates speech waveforms from acoustic features \cite{parallel_wavegan}. 
The detailed model structure and training conditions of these two models were the same as those in \cite{yamamoto2020pwgvuvd}.
By inputting the predicted phrase breaks along with the input phoneme sequence, the TTS system generated target speech waveforms.

As a database, we used the same corpus as that used in the phrase break prediction experiments.
The speech corpus was recorded by a single Japanese professional speaker.
The speech signals were sampled at 24 kHz, and each sample was quantized by 16 bits.
The total amount of the training data size was 118.9 hours.
Note that we used the same test set as that used in phrase break prediction, which was not included in the training set, for evaluations.

\subsubsection{Subjective evaluation setup}
We performed an MOS test and AB preference test for the subjective evaluations.
In the MOS test, 29 native Japanese speakers were asked to make quality judgments about the synthesized speech samples using the following five possible responses: 1 = Bad; 2 = Poor; 3 = Fair; 4 = Good; 5 = Excellent.
In the AB preference test, each test case consisted of two audio samples, A and B, where A and B correspond to the synthesized speech samples of different systems.
The same subjects of the MOS test were asked to choose the audio sample A or B that was more natural than the other, or state whether A and B were the same in terms of the naturalness of the prosody.

For the MOS test, we compared the following seven systems: (1) \textbf{Reference (Natural)}: recorded speech in the test set; (2) \textbf{Reference (TTS)}: synthesized speech from the test set; (3) \textbf{Rule-based}: a rule-based method that inserts phrase breaks only after punctuation, and some methods used in the objective evaluations; (4) \textbf{BiLSTM (Tokens)}, (5) \textbf{BiLSTM (Features)}, (6) \textbf{BERT}, and (7) \textbf{BiLSTM (Features) + BERT}.
For the AB preference test, we compared the following six system pairs; (3) Rule-based and (4) BiLSTM (Tokens); (4) BiLSTM (Tokens) and (5) BiLSTM (Features); (5) BiLSTM (Features) and (6) BERT; (6) BERT and (7) BiLSTM (Features) + BERT; (6) BERT and (2) Reference (TTS); and (7) BiLSTM (Features) + BERT and (2) Reference (TTS).
Audio samples for each system are available on the website\footnote{\url{https://matasuke.github.io/demos/pbp_bert}}.

For the MOS test, 30 utterances were randomly selected from the test set and then synthesized using the above seven models; in total, 210 utterances were used.
For the AB preference test, 30 utterances were randomly selected from the test set and then synthesized using the above six models; in total, 360 utterances were used.

\subsection{Results and analysis}
Table \ref{tab:mos_evaluation_result} shows the MOS test results for the different systems.
The experimental results of the MOS test showed that the proposed method achieved almost the same quality as Reference (TTS), which synthesized test data, and indicated a significant improvement in the naturalness of the prosody compared with conventional methods.
Comparing BiLSTM (Features) + BERT with the conventional methods, BiLSTM (Tokens) and BiLSTM (Features), the proposed method achieved 4.39 MOS points and improved by 0.59 points and 0.34 points, respectively.
Additionally, the proposed method improved by 0.13 MOS points compared with BERT.
This indicates that using both explicit features from BiLSTM and implicit features from BERT was effective.

Table \ref{tab:ab_test_result} shows the AB preference test results with respect to different systems:
n.s. denotes no significant difference, and *** denotes statistically significant difference at the 1\% level as a result of a one-tailed binomial test.
The experimental results of the AB preference test showed significant differences at the 1\% level for five pairs, except for the results between BiLSTM (Features) + BERT and Reference (TTS).
For the pair between BiLSTM (Features) + BERT and Reference (TTS), the preferences for each system were competitive, and there was no significant difference between them.
Hence, we assume that the proposed method achieved almost the same quality as the synthesized speech samples using the reference text.

\section{Conclusions}
In this paper, inspired by the great success of BERT in many NLP tasks, we presented a novel method using both labeled data and unlabeled data for phrase break prediction in a Japanese TTS front-end.
The objective evaluation results showed that the BiLSTM and BERT-based proposed method significantly improved the performance of phrase break prediction compared with conventional methods and the model based only on BERT.
Moreover, subjective evaluation results verified that the proposed method achieved almost the same quality as the synthesized speech samples using reference text in terms of the naturalness of the prosody.
In the future, we will aim to further improve the quality of synthesized speech samples in terms of the naturalness of the prosody by modeling the duration of phrase breaks.

\section{Acknowledgements}
This work was supported by Clova Voice, NAVER Corp., Seongnam, Korea.

\clearpage
\bibliographystyle{IEEEtran}
\bibliography{mybib}

\end{document}